\documentclass[conference]{IEEEtran}
\IEEEoverridecommandlockouts
\usepackage{cite}
\usepackage{amsmath,amssymb,amsfonts}
\usepackage{algorithmic}
\usepackage{graphicx}
\usepackage{textcomp}
\usepackage{xcolor}
\usepackage{url}
\usepackage[hidelinks]{hyperref}

\def\BibTeX{{\rm B\kern-.05em{\sc i\kern-.025em b}\kern-.08em
    T\kern-.1667em\lower.7ex\hbox{E}\kern-.125emX}}
\begin{document}

\title{Probabilistic Stellar Age Estimation for Gaia XP Stars with NGBoost\\
\thanks{This work was supported by the Postdoctoral Research Fund of Inner Mongolia University under Grant 10000-A260015/063, the Research Start-up Fund for Young Academic Talents of Inner Mongolia University under Grant 10000-A260015/562, the National Natural Science Foundation of China under grant No. 12588202, the Strategic Priority Research Program of the Chinese Academy of Sciences under grant No. XDB1160103, the National Key R\&D Program of China under grant Nos. 2024YFA1611903 and 2023YFE0107800.}
}

\author{\IEEEauthorblockN{Xiaokun Hou*}
\IEEEauthorblockA{\textit{Institute of Astronomy and Physics} \\
\textit{Inner Mongolia University}\\
Hohhot, China \\
houxiaokun@imu.edu.cn}
*Corresponding author
~\\
\and
\IEEEauthorblockN{Wenbo Wu*}
\IEEEauthorblockA{\textit{Institute of Astronomy and Physics} \\
\textit{Inner Mongolia University}\\
Hohhot, China \\
wbwu@imu.edu.cn}
*Corresponding author
~\\
\and
\IEEEauthorblockN{Gang Zhao}
\IEEEauthorblockA{\textit{Institute of Astronomy and Physics} \\
\textit{Inner Mongolia University}\\
Hohhot, China}
\IEEEauthorblockA{\textit{National Astronomical Observatories} \\
\textit{Chinese Academy of Sciences}\\
Beijing, China \\
gzhao@nao.cas.cn}
~\\
\and
\IEEEauthorblockN{Haining Li}
\IEEEauthorblockA{\textit{Institute of Astronomy and Physics} \\
\textit{Inner Mongolia University}\\
Hohhot, China}
\IEEEauthorblockA{\textit{National Astronomical Observatories} \\
\textit{Chinese Academy of Sciences}\\
Beijing, China \\
lhn@nao.cas.cn}
~\\
\and
\IEEEauthorblockN{Jingkun Zhao}
\IEEEauthorblockA{\textit{Institute of Astronomy and Physics} \\
\textit{Inner Mongolia University}\\
Hohhot, China}
\IEEEauthorblockA{\textit{National Astronomical Observatories} \\
\textit{Chinese Academy of Sciences}\\
Beijing, China \\
zjk@bao.ac.cn}
}

\maketitle
\begin{abstract}
Stellar age is a fundamental quantity for Galactic archaeology, but reliable age estimation for large stellar samples remains challenging. In this work, we develop an uncertainty-aware NGBoost framework for stellar age estimation using Gaia XP-derived atmospheric parameters and chemical abundances. Different from the standard NGBoost model, we modify the loss function by incorporating the uncertainties of the training age labels. We further use a Monte Carlo strategy to quantify the influence of input-feature uncertainties on the predicted ages. The resulting model provides age estimates together with uncertainty estimates. Applying this framework to Gaia XP stars, we construct a stellar age catalog containing 15,175,107 stars. 
\end{abstract}


\section{Introduction}
Stellar age is one of the most fundamental quantities in Galactic Astronomy. Together with chemical abundances and phase-space information, stellar ages provide the time dimension required to reconstruct the formation and evolution history of the Milky Way. However, stellar age is notoriously difficult to measure. Unlike position, velocity, or brightness, age is not directly observable and must be inferred from stellar evolution models or empirical correlations. Asteroseismology provides one of the most powerful approaches for determining stellar ages, because stellar oscillations can constrain stellar masses and radii, which are closely related to evolutionary stage and age \cite{b1}. Space missions such as Kepler \cite{b2} and Tess \cite{b3} have provided high-precision, nearly continuous photometric time series that make such measurements possible. Nevertheless, asteroseismic age estimation requires long-duration light curves for individual stars, leaving only limited age samples for large-scale Galactic studies.

Machine-learning methods offer a promising way to extend precise age information from small asteroseismic samples to much larger surveys. A common strategy is to use asteroseismic ages as training labels and spectra or spectroscopically derived abundances as input features, allowing models to learn age-sensitive information from spectroscopic data. Several studies have inferred stellar ages from APOGEE \cite{b4} high-resolution spectra. For example, the astroNN framework \cite{b5}, originally developed to extract atmospheric parameters and chemical abundances from spectra, has been applied to stellar age estimation \cite{b6}. Anders et al. \cite{b7} trained an XGBoost \cite{b8} model using APOGEE chemical abundances and Kepler astroseismic ages, obtaining precise age estimates and revealing spatial, chemical, and kinematic trends with age in the Galactic disk. Despite these successes, spectroscopic-age catalogs remain limited by survey footprints, target selection, magnitude limits, and other selection effects. These limitations are critical for studies of the global Milky Way structure, where spatial coverage and completeness are essential. Therefore, extending age estimation from high-resolution spectroscopic surveys to large-scale photometric data sets, is essential for exploiting the full potential of modern Galactic surveys.

Gaia \cite{b9} provides an unprecedented opportunity in this direction. In addition to all-sky astrometry, Gaia DR3 provides low-resolution BP/RP spectrophotometry, commonly referred to as XP spectra, for about 220 million sources. Although Gaia XP spectra have much lower spectral resolution than APOGEE, their enormous sample size and broad sky coverage make them uniquely valuable for Galactic studies. The XP spectra are released as continuous spectra represented by basis-function coefficients, from which sampled BP/RP spectra can be reconstructed. Recent machine-learning studies have shown that Gaia XP spectra can be used to infer stellar atmospheric parameters, as well as chemical abundances such as carbon, nitrogen, and $\alpha$-elements \cite{b10}. Because stellar parameters and chemical abundances encode information from both stellar evolution and Galactic chemical enrichment, they can serve as useful age indicators. Gaia XP-derived parameters and abundances therefore provide a natural basis for large-scale stellar age estimation beyond high-resolution spectroscopic surveys.

In this work, we develop an NGBoost-based probabilistic framework for stellar age estimation using stellar parameters and abundances derived from Gaia XP spectra. Unlike XGBoost models that provide only point estimates, our model predicts a full age distribution. It also accounts for uncertainties in both the training ages and the input parameters and abundances. The paper is organized as follows. Section \ref{secdata} describes the data sets used in this work. Section \ref{secmethod} presents the NGBoost-based probabilistic model and explains how label uncertainties and input errors are incorporated. Section \ref{secresults} presents the resulting stellar age catalog and evaluates the model performance. Finally, Section \ref{secconclusion} summarizes the main conclusions.

\section{Data}\label{secdata}
The NGBoost-based age model developed in this work requires stellar parameters and chemical abundances as input features. Specifically, the input vector includes effective temperature ($T_{\rm{eff}}$), surface gravity ($\rm{log}\,\textsl{g}$), metallicity ([Fe/H]), and the carbon, nitrogen, and $\alpha$-element abundances ([C/Fe], [N/Fe], and [$\alpha$/M]). The uncertainties of these input features are used to quantify how measurement errors propagate into the predicted stellar ages. In addition, stellar ages and their uncertainties are required as training labels. Therefore, the data used in this work are mainly drawn from two catalogs. The first is the Gaia XP-based catalog constructed by Fallows and Sanders \cite{b10}, which provides stellar parameters, chemical abundances, and their corresponding uncertainties using uncertain neural networks. The second is the astroNN value-added catalog for APOGEE DR17, which provides stellar ages and the corresponding age uncertainties. 

\subsection{The Reference Set}
The reference set is constructed by cross-matching the two catalogs described above. Before cross-matching, duplicate entries are removed from both catalogs. In the astroNN catalog, we retain the entry with the smallest \texttt{age\_total\_error} for each duplicate source and remove stars without age measurements. In the Gaia XP value-added catalog, we retain the entry with the smallest \texttt{teff\_unc} for each duplicate source. The two catalogs are then cross-matched using Gaia \texttt{source\_id} as the common identifier, resulting in 489,913 matched sources.

\begin{figure*}[htbp]
\centering{\includegraphics[width=1.0\linewidth]{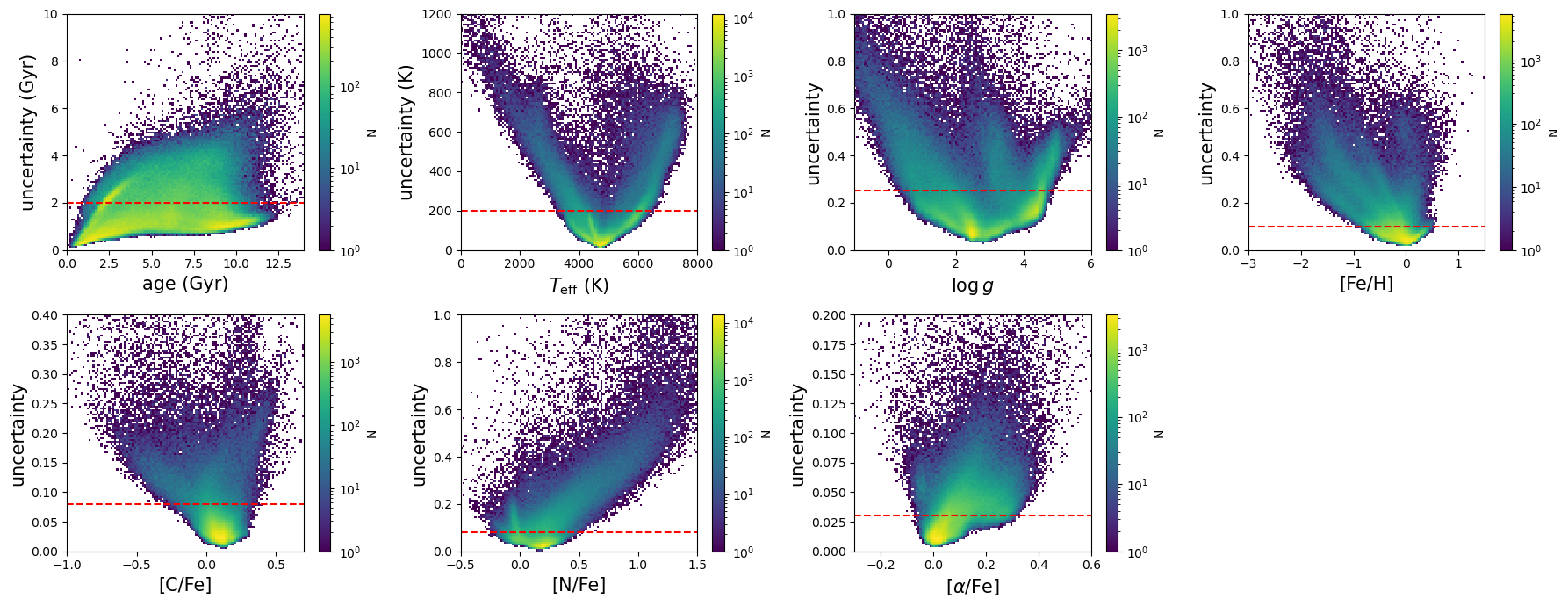}}
\caption{Quality assessment of the cross-matched sample. Each panel shows the distribution of stellar age, atmospheric parameter, or chemical abundance versus its corresponding uncertainty. The color scale indicates the number density, and the red dashed lines mark the adopted uncertainty cuts.}
\label{fig1}
\end{figure*}

\begin{figure}[htbp]
\centering{\includegraphics[width=0.97\linewidth]{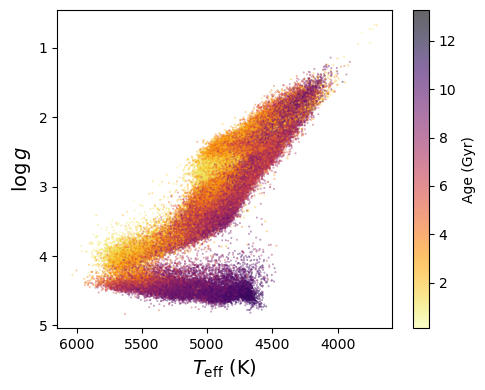}}
\caption{Kiel diagram of the reference stars, color-coded by stellar age.}
\label{fig2}
\end{figure}

We further inspect the quality of the matched sources and find that some stars have poorly constrained parameters, abundances, or ages, as indicated by large uncertainties. Fig.~\ref{fig1} shows the distributions of stellar ages, atmospheric parameters, and chemical abundances against their corresponding uncertainties. To ensure the reliability of the reference set, we apply additional quality cuts to the matched sample, as summarized in Table \ref{tab1}. We also remove sources with \texttt{RUWE} larger than 1.4, because their astrometric measurements are likely unreliable \cite{b11}. After these cuts, 135,569 sources remain in the reference set. The parameter and abundance ranges of the reference set are listed in Table \ref{tab2}, which defines the applicability domain of the trained model. The Kiel diagram of the reference set is shown in Fig.~\ref{fig2}, where the stars are color-coded by age. The reference set is then randomly divided into training, validation, and test sets with a ratio of 70\%, 10\%, and 20\%, respectively.

\begin{table}[htbp]
\caption{Quality cuts applied to the cross-matched sample.}
\begin{center}
\begin{tabular}{|c|c|}
\hline
\textbf{Quantity} & \textbf{Selection criterion} \\
\hline
Age uncertainty & $< 2.0$ (Gyr) \\
$T_{\rm eff}$ uncertainty & $< 200$ (K) \\
$\log g$ uncertainty & $< 0.25$ \\
{[Fe/H]} uncertainty & $< 0.10$  \\
{[C/Fe]} uncertainty & $< 0.08$  \\
{[N/Fe]} uncertainty & $< 0.08$  \\
{[$\alpha$/M]} uncertainty & $< 0.03$  \\
RUWE & $< 1.4$ \\
\hline
\end{tabular}
\label{tab1}
\end{center}
\end{table}

\begin{table}[htbp]
\caption{Parameter and abundance ranges of the final reference set.}
\label{tab:parameter_ranges}
\centering
\begin{tabular}{|c|c|c|}
\hline
\textbf{parameter/abundance} & \textbf{Minimum} & \textbf{Maximum} \\
\hline
$T_{\rm eff}$ & 3699.92 (K) & 6032.32 (K) \\
$\log g$  & 0.66 & 4.83 \\
{[Fe/H]}  & -0.76 & 0.53 \\
{[C/Fe]}  & -0.20 & 0.34 \\
{[N/Fe]}  & -0.20 & 0.51 \\
{[$\alpha$/M]}  & -0.05 & 0.33 \\
\hline
\end{tabular}
\label{tab2}
\end{table}

\section{Method}\label{secmethod}
\subsection{Overview of NGBoost}
Unlike point-estimation models such as XGBoost, NGBoost \cite{b12} represents the prediction as a probability distribution. For continuous regression problems, this distribution is commonly specified as a Gaussian distribution. In this case, NGBoost does not directly output a single value, but instead predicts two parameters of the Gaussian distribution, the mean ($\mu_i$) and the standard deviation ($\sigma_i$). These two parameters are denoted as $\theta_i$ ($\theta_i=(\mu_i,\sigma_i)$), where the subscript $i$ refers to the $i$-th input sample. 

Since NGBoost outputs a Gaussian distribution for each input vector $\mathbf{x}_i$, the target value $y_i$ can be evaluated by its conditional probability density given $\mathbf{x}_i$. Specifically, the model first maps $\mathbf{x}_i$ to the distribution parameters $\theta_i$, and these parameters define the Gaussian probability density ($p(y_i|\mathbf{x}_i;\theta_i)$). For a training set with N samples, the likelihood is the product of the predicted probability densities of all labels:
\begin{equation}
    \prod_{i=1}^{N}  p(y_i|\theta_i)
\end{equation}
Maximizing this likelihood is equivalent to minimizing the negative log-likelihood, which can be written as:
\begin{equation}
    L=-\sum_{i=1}^{N} log\, p(y_i|\theta_i)
\end{equation}
Under the Gaussian assumption, this loss function becomes:
\begin{equation}
    L=-\sum_{i=1}^{N}  [0.5\times log(2\pi\sigma_i^2)+\frac{(y_i-\mu_i)^2}{2\sigma_i^2}]
\end{equation}
By minimizing this loss function, NGBoost learns the mapping from each input vector $\mathbf{x}_i$ to the distribution parameters $\theta_i$. Therefore, instead of learning $\mathbf{x}_i$ to $y_i$ as in point-estimation models, NGBoost learns $\mathbf{x}_i$ to $\theta_i$, which defines a full predictive distribution.

\subsection{Uncertainty-aware NGBoost Model}
Although NGBoost provides a predictive distribution, the standard NGBoost loss function treats the training labels as exact values. This is not fully appropriate for stellar age estimation, because the age labels used for training have non-negligible uncertainties. To include this effect, we modify the likelihood by distinguishing between intrinsic quantities and observed quantities.

For each star, we denote the intrinsic age as $A_t$ and the intrinsic input vector as $\mathbf{x}_t$. The input vector $\mathbf{x}_t$ consists of the intrinsic effective temperature ($T_{\rm{eff}}$), surface gravity ($\rm{log}\,\textsl{g}$), metallicity ([Fe/H]), and the carbon, nitrogen, and $\alpha$-element abundances ([C/Fe], [N/Fe], and [$\alpha$/M]). The physical relation to be learned is therefore the mapping from $\mathbf{x}_t$ to $A_t$. However, neither $A_t$ nor $\mathbf{x}_t$ can be obtained. The astroNN catalog provides an observed age $A_{obs}$ with uncertainty $\sigma_{A,obs}$, while the Gaia XP value-added catalog provides observed input vector $\mathbf{x}_{obs}$ with uncertainties $\sigma_{\mathbf{x},{obs}}$.

For a given intrinsic input vector $\mathbf{x}_t$, NGBoost predicts a Gaussian distribution for the intrinsic age:
\begin{equation}
      A_t|\mathbf{x}_t \sim \mathcal{N}(\mu,\sigma^2)
\label{eq4}
\end{equation}
where $\mu$ and $\sigma$ are the model-predicted mean and standard deviation. During training, however, the intrinsic age is unavailable, and only the observed age can be used as the label. We therefore model the observed age as intrinsic age plus a measurement error:
\begin{equation}
      A_{obs}=A_t+\epsilon_A
\end{equation}
where the age measurement error follows:
\begin{equation}
    \epsilon_A \sim \mathcal{N}(0,\sigma_{A,obs}^2)
\end{equation}
This gives:
\begin{equation}
    A_{obs}|A_t \sim \mathcal{N}(A_t,\sigma_{A,obs}^2)
\label{eq7}
\end{equation}
Since $A_t$ is unavailable, the conditional density of $A_{obs}$ given $\mathbf{x}_t$ is obtained by marginalizing over the intrinsic age:
\begin{equation}
   p(A_{obs}|\mathbf{x}_t)=\int p(A_{obs}|A_t)p(A_t|\mathbf{x}_t)dA_t
\end{equation}
Because both terms are Gaussian, the marginal distribution is also Gaussian:
\begin{equation}
       A_{obs}|\mathbf{x}_t \sim \mathcal{N}(\mu,\sigma^2+\sigma_{A,obs}^2)
\label{eq9}
\end{equation}
With this formulation, the training label is changed from the inaccessible intrinsic age to the observed age. At the same time, the uncertainty of the age label is incorporated into the loss function through the total variance ($\sigma^2+\sigma_{A,obs}^2$).

Similarly, the intrinsic input vector is also unavailable. During model training, we use the $\mathbf{x}_{ obs}$ as an approximation of $\mathbf{x}_t$, because the conditional probability ($p(\mathbf{x}_t|\mathbf{x}_{obs})$) is difficult to determine. With this approximation, the distribution in \eqref{eq9} can be written as:
\begin{equation}
       A_{obs}|\mathbf{x}_{obs} \sim \mathcal{N}(\mu,\sigma^2+\sigma_{A,obs}^2)
\label{eq10}
\end{equation}
The uncertainties of the input features are propagated during prediction, including both the test set and stars outside the reference set. For each star, we treat the observed input vector as a multivariate Gaussian distribution with a diagonal covariance matrix. The diagonal elements are given by the squared uncertainties of the corresponding input features, including ($T_{\rm{eff}}$), ($\rm{log}\,\textsl{g}$), [Fe/H], [C/Fe], [N/Fe], and [$\alpha$/M]. We then draw 200 Monte Carlo realizations of the input vector. For each realization, the trained NGBoost model predicts a Gaussian age distribution with mean ($\mu_j$) and standard deviation ($\sigma_j$). The final predicted age is taken as the average of the Monte Carlo mean predictions. The standard deviation of the Monte Carlo means is used to quantify the propagated age uncertainty caused by input-feature errors. We estimate the final model uncertainty by taking the root mean square of $\sigma_j$ over all Monte Carlo realizations. Finally, the total age uncertainty is obtained by adding the propagated uncertainty and the model uncertainty in quadrature.

\section{Results and Discussion}\label{secresults}
\begin{figure*}[htbp]
\centering{\includegraphics[width=1.0\linewidth]{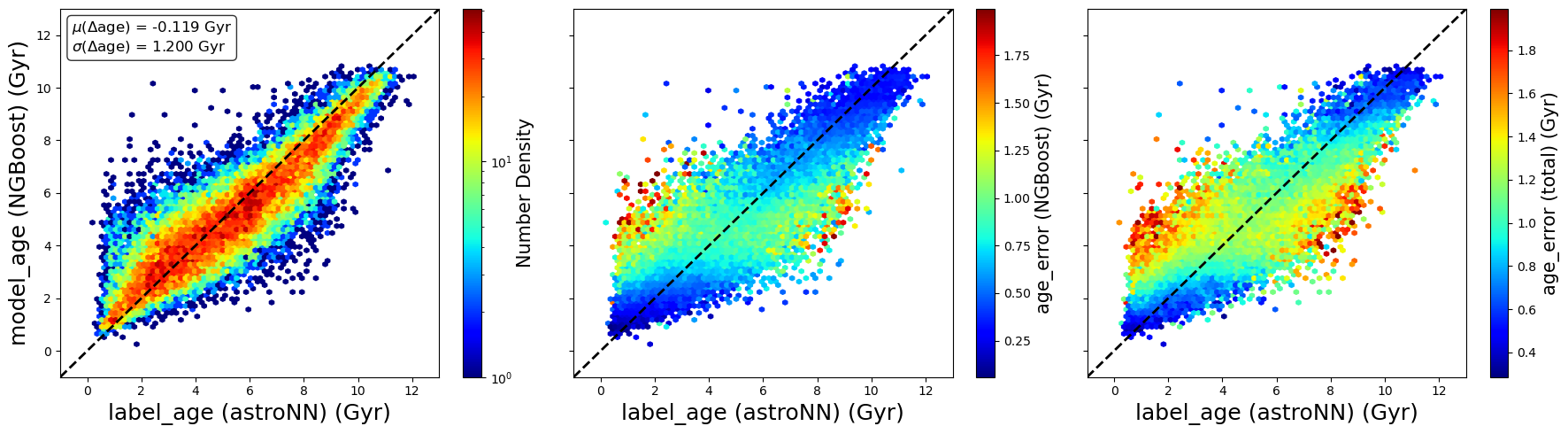}}
\caption{Performance of the NGBoost-based age model on the test set. The predicted ages are compared with the astroNN reference ages. The left panel is color-coded by number density, while the middle and right panels are color-coded by the model uncertainty and the total uncertainty, respectively. The black dashed line indicates the one-to-one relation. The mean and standard deviation of the age residuals are shown in the left panel.
}
\label{fig3}
\end{figure*}

The performance of the trained model is first evaluated on the test set by comparing the NGBoost-predicted ages with the reference ages from astroNN. As shown in Fig. \ref{fig3}, the left panel is color-coded by number density and shows that the main population of test stars is concentrated along the one-to-one relation over a wide age range. The mean age residual, defined as the difference between the predicted age and the reference age, is -0.119 Gyr, with a scatter of 1.200 Gyr, indicating that the model introduces only a small systematic offset. The middle and right panels further show the same comparison color-coded by the NGBoost model uncertainty and the total age uncertainty, respectively. Stars with larger deviations from the one-to-one relation generally have larger predicted uncertainties, while stars closer to the one-to-one line tend to have smaller uncertainties. This behaviour indicates that the uncertainty estimates are informative and can be used to assess the reliability of individual age predictions. Compared with point-estimation methods such as XGBoost, the probabilistic framework adopted here provides not only stellar age estimates but also corresponding uncertainties, allowing more reliable age samples to be selected by applying uncertainty cuts.

After validating the model on the test set, we apply the trained model to the Gaia XP value-added catalog of Fallows and Sanders \cite{b10} to construct a stellar age catalog. We first select stars that satisfy the uncertainty cuts listed in Table \ref{tab1} and fall within the model applicability ranges given in Table \ref{tab2}. This selection yields 15,175,107 stars. We then input their atmospheric parameters and chemical abundances into the trained model and obtain probabilistic age estimates for these stars. The resulting catalog includes the predicted age, model uncertainty, propagated uncertainty, and total age uncertainty, and is made available online \footnote{\url{https://github.com/tubage886/Age-for-Gaia-XP-Stars}}.

\begin{figure*}[htbp]
\centering{\includegraphics[width=1\linewidth]{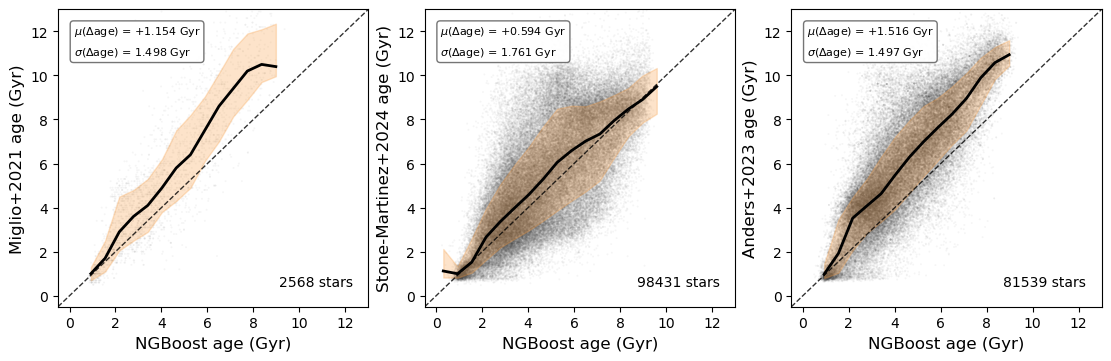}}
\caption{Comparison between our ages and literature ages. The three panels show comparisons with the ages from Miglio et al. \cite{b13}, Stone-Martinez et al. \cite{b14}, and Anders et al. \cite{b7}, respectively. Grey points represent individual stars. The black dashed line indicates the one-to-one relation. The solid lines and orange shaded regions show the running median and 1$\sigma$ quantiles, respectively. The number of stars used in each comparison is marked in the lower-right corner, and the mean and standard deviation of the age differences are shown in the upper-left corner.}
\label{fig4}
\end{figure*}

To further assess the reliability of the resulting age catalog, we compare our ages with stellar ages reported in previous studies. Miglio et al. \cite{b13} estimated ages for red giants in the Kepler field using asteroseismology, while Stone-Martinez et al. \cite{b14} and Anders et al. \cite{b7} derived ages for APOGEE stars using neural-network and XGBoost-based methods, respectively. We cross-match our catalog with these three catalogs and retain only stars whose total age uncertainty in our catalog is smaller than 1.4 Gyr. The comparison results are shown in Fig. \ref{fig4}. In Fig. 4, the three panels compare the NGBoost ages with the literature ages from Miglio et al. \cite{b13}, Stone-Martinez et al. \cite{b14}, and Anders et al. \cite{b7}, respectively. Individual stars are shown as grey points. The number of stars used in each comparison is marked in the lower-right corner of each panel, while the mean and standard deviation of the age difference are shown in the upper-left corner. The black dashed line indicates the one-to-one relation. The solid lines and shaded bands in each panel delineate the running median and 1$\sigma$ quantiles, respectively. Overall, the NGBoost ages show good agreement with the literature ages. In the comparisons with Miglio et al. \cite{b13} and Anders et al. \cite{b7}, the positive mean values of $\Delta{\rm Age}$ indicate that our NGBoost ages are systematically younger than the corresponding literature estimates. This offset is likely dominated by the zero-point difference between the age scale of our training labels and those of the two literature catalogs. Indeed, a direct comparison between astroNN ages and the Miglio et al. \cite{b13} and Anders et al. \cite{b7} ages shows the same tendency for astroNN to give younger age estimates. The observed offsets therefore mainly reflect differences in the underlying training age scale, which are subsequently propagated into the NGBoost predictions. However, the relatively narrow 1$\sigma$ bands suggest that the scatter of the age difference is limited. In the comparison with Stone-Martinez et al. \cite{b14}, the age difference shows a larger scatter, but no strong systematic trend is seen over most of the age range. The standard deviations of the age differences are about $1.5-1.8$ Gyr in the three comparisons, indicating that our age estimates are broadly consistent with independent age catalogs derived using asteroseismology, neural networks, and XGBoost-based methods.

It is also useful to compare our work with the recent age catalog of Almannaei et al. \cite{b15}. Although both studies use the stellar atmospheric parameters and chemical abundances derived from the Fallows and Sanders Gaia XP catalog \cite{b10} as inputs for stellar age inference, they differ in two important aspects. First, the stellar parameter and abundance ranges over which the models are applied are different. Almannaei et al. \cite{b15} focused on giant stars and massive red-clump stars, selecting stars within $4000<T_{\rm{eff}}<5400$ (K) and $1.5<\rm{log}\,\textsl{g}<3.5$. To reduce contamination from low-mass red-clump stars, they further excluded stars with $\log(T_{\rm{eff}})>3.675$ and $\rm{log}\,\textsl{g}<2.6$. In contrast, our catalog includes both giant and dwarf stars within the parameter ranges defined in Table \ref{tab2}, and no such additional red-clump-related selection cut is applied. Therefore, our age catalog provides broader stellar coverage and is less affected by this specific selection criterion. Second, the treatment of uncertainty is different. Almannaei et al. \cite{b15} used XGBoost to estimate stellar ages, which provides point predictions and does not explicitly account for the effects of training-label uncertainties and input-feature uncertainties on the inferred ages. As a result, the resulting ages do not have individual uncertainty estimates, making it difficult to assess the reliability of each predicted age and limiting their use in studies that require well-controlled age precision. In contrast, our NGBoost-based framework explicitly incorporates the uncertainties of the training age labels, propagates the observational uncertainties of the input parameters and abundances, and also provides a model uncertainty for each star. This probabilistic treatment allows the quality of the inferred ages to be controlled through the predicted uncertainties, which is particularly important for constructing reliable age-selected samples for Galactic archaeology.

\section{Conclusion}\label{secconclusion}
In this work, we developed an uncertainty-aware NGBoost framework for stellar age estimation. Different from the standard NGBoost model, we modified the loss function by incorporating the uncertainties of the training age labels, so that the observed ages and their errors can be directly used during model training. In addition, we used a Monte Carlo strategy to quantify the effects of input-feature uncertainties on the predicted ages. This allows the model to provide not only age estimates but also individual uncertainty estimates that include both model uncertainty and propagated uncertainty. 

Using the trained model, we constructed a Gaia XP age catalog comprising 15,175,107 stars whose atmospheric parameters and chemical abundances fall within the applicable range of the model. The catalog provides \texttt{source\_id}, \texttt{predicted\_age}, \texttt{model\_uncertainty}, \texttt{propagated\_uncertainty}, and \texttt{total\_uncertainty}, and is made publicly available online. Its consistency with literature age estimates further supports the reliability of the inferred stellar ages.

Overall, by modifying the NGBoost loss function and using a Monte Carlo strategy, this work enables the standard NGBoost framework to effectively incorporate the effects of age-label uncertainties and input-feature uncertainties. With this uncertainty-aware framework, we derive probabilistic ages for more than 15 million Gaia XP stars. The resulting catalog offers valuable data support for future studies of the formation and evolution history of the Milky Way.

\section*{Acknowledgment}
This work has made use of data from the European Space Agency (ESA) mission Gaia, processed by the Gaia Data Processing and Analysis Consortium (DPAC). Funding for the DPAC has been provided by national institutions, in particular the institutions participating in the Gaia Multilateral Agreement. This work also made use of data from the Apache Point Observatory Galactic Evolution Experiment (APOGEE), which is part of the Sloan Digital Sky Survey IV (SDSS-IV). Funding for SDSS-IV has been provided by the Alfred P. Sloan Foundation, the U.S. Department of Energy Office of Science, and the Participating Institutions. The authors also acknowledge the public Gaia XP value-added catalog of Fallows and Sanders and the astroNN value-added catalog used in this study.

\vspace{12pt}


\begin{thebibliography}{00}
\bibitem{b1}
W. J. Chaplin and A. Miglio, ``Asteroseismology of solar-type and red-giant stars,'' \emph{Annu. Rev. Astron. Astrophys.}, vol. 51, no. 1, pp. 353--392, Aug. 2013.

\bibitem{b2} D. G. Koch et al., ``Kepler mission design, realized photometric performance, and early science,'' \emph{Astrophys. J. Lett.}, vol. 713, no. 2, pp. L79--L86, Apr. 2010.

\bibitem{b3}
G. R. Ricker et al., ``Transiting Exoplanet Survey Satellite (TESS),'' \emph{J. Astron. Telesc. Instrum. Syst.}, vol. 1, Art. no. 014003, Jan. 2015.

\bibitem{b4} S. R. Majewski et al., ``The Apache Point Observatory Galactic Evolution Experiment (APOGEE),'' \emph{Astron. J.}, vol. 154, no. 3, Art. no. 94, Sep. 2017.

\bibitem{b5} H. W. Leung and J. Bovy, ``Deep learning of multi-element abundances from high-resolution spectroscopic data,'' \emph{Mon. Not. R. Astron. Soc.}, vol. 483, no. 3, pp. 3255--3277, Mar. 2019,.

\bibitem{b6} J. T. Mackereth et al., ``Dynamical heating across the Milky Way disc using APOGEE and Gaia,'' \emph{Mon. Not. R. Astron. Soc.}, vol. 489, no. 1, pp. 176--195, Oct. 2019.

\bibitem{b7} F. Anders et al., ``Spectroscopic age estimates for APOGEE red-giant stars: Precise spatial and kinematic trends with age in the Galactic disc,'' \emph{Astron. Astrophys.}, vol. 678, Art. no. A158, Oct. 2023.

\bibitem{b8}
T. Chen and C. Guestrin, ``XGBoost: A scalable tree boosting system,'' in \emph{Proc. 22nd ACM SIGKDD Int. Conf. Knowl. Discov. Data Min.}, San Francisco, CA, USA, 2016, pp. 785--794.

\bibitem{b9} Gaia Collaboration, T. Prusti et al., ``The Gaia mission,'' \emph{Astron. Astrophys.}, vol. 595, Art. no. A1, Nov. 2016.

\bibitem{b10} C. P. Fallows and J. L. Sanders, ``Stellar atmospheric parameters from Gaia BP/RP spectra using uncertain neural networks,'' \emph{Mon. Not. R. Astron. Soc.}, vol. 531, no. 1, pp. 2126--2147, Jun. 2024.

\bibitem{b11} L. Lindegren et al., ``Gaia Early Data Release 3. Parallax bias versus magnitude, colour, and position,'' \emph{Astron. Astrophys.}, vol. 649, Art. no. A4, May 2021.

\bibitem{b12}T. Duan et al., ``NGBoost: Natural gradient boosting for probabilistic prediction,'' in \emph{Proc. 37th Int. Conf. Mach. Learn.}, 2020, pp. 2690--2700.

\bibitem{b13} A. Miglio et al., ``Age dissection of the Milky Way discs: Red giants in the Kepler field,'' \emph{Astron. Astrophys.}, vol. 645, Art. no. A85, Jan. 2021.

\bibitem{b14} A. Stone-Martinez, J. A. Holtzman, J. Imig, C. Nitschelm, K. G. Stassun, and J. R. Brownstein, ``Spectroscopic distance, mass, and age estimations for APOGEE DR17,'' \emph{Astron. J.}, vol. 167, no. 2, Art. no. 73, Jan. 2024.

\bibitem{b15}A. S. Almannaei et al., ``Towards Galactic archaeology with inferred ages of giant stars from Gaia spectra,'' \emph{Mon. Not. Roy. Astron. Soc.}, vol. 546, no. 1, Art. no. staf2252, Feb. 2026.



\end{thebibliography}
\end{document}